# Assessing Vibrational Frequencies of CO Adsorbed on Cerium Oxide Surfaces Using SCAN and r²SCAN Functionals

Alexander Contreras-Payares,[1,2] Pablo G. Lustemberg,[1] M. Verónica Ganduglia-Pirovano[1*]

[1]*Institute of Catalysis and Petrochemistry, ICP, Spanish National Research Council, CSIC, 28049 Madrid, Spain*

[2]*PhD Theoretical Chemistry and Computational Modelling Doctoral School, Universidad Autónoma de Madrid, C/Francisco Tomás y Valiente 2, 28049 Ciudad Universitaria de Cantoblanco, Madrid, Spain*

**ABSTRACT**

The vibrational frequency of carbon monoxide (CO) adsorbed on ceria-based catalysts serves as a sensitive probe for identifying exposed surface facets, provided that experimental reference data on well-defined single-crystal surfaces and reliable theoretical assignments are available. Previous studies have shown that the hybrid DFT approach using the HSE06 functional yields good agreement with experimental observations, whereas the generalized gradient approximation (GGA) with PBE+U does not. In this work, we assess the performance of different exchange-correlation functionals by comparing the meta-GGA functionals SCAN and r²SCAN meta-GGA functionals with HSE06 in predicting CO vibrational frequencies on cerium oxide surfaces. The meta-GGA functionals offer no significant improvement for oxidized $CeO_2(111)$ and $CeO_2(110)$ surfaces and fail to localize excess charge on the reduced surfaces. Adding a Hubbard U term improves charge localization, but the predicted vibrational frequencies still fall short of HSE06 accuracy. These limitations are attributed to the meta-GGA's inability to adequately capture facet- and configuration-specific donation and back-donation effects, which influence the C—O bond length and CO force constant upon adsorption. Despite the higher computational cost when used with plane-wave basis sets, hybrid DFT remains essential for accurate interpretation of experimental results.

*Corresponding authors: vgp@icp.csic.es

I.  **INTRODUCTION**

Cerium dioxide (CeO$_2$), commonly known as ceria, is a key material in heterogeneous catalysis, serving both as a catalyst and as a support.[1] Beyond catalysis, it is also used in solid-oxide fuel cells and is gaining attention in biological applications.[2,3] In catalytic systems, ceria is typically employed in powder form, consisting of nano- or microscale particles with distinct crystallographic facets. These surface variations significantly influence chemical reactivity.[4–10] Additionally, the formation of oxygen vacancies, which is facet dependent,[11] and structural changes under reaction conditions are critical factors affecting performance.[12–17]

To accurately analyze the chemical behavior of ceria nanoparticles, reliable methods for identifying exposed surface facets are essential. Among various experimental approaches, the use of carbon monoxide (CO) as a probe molecule in infrared (IR) vibrational spectroscopy—particularly via the surface-ligand IR (SLIR) technique[18] —has proven highly effective.[19–21] While numerous studies have investigated CO adsorption on ceria powders [22–26] comprehensive characterization of technologically relevant ceria materials still requires experimental and computational reference data for well-defined oxidized and reduced surfaces. Theoretical methods, especially those based on density functional theory (DFT) using the generalized gradient approximation (GGA) for exchange and correlation,[27] have faced persistent difficulty modeling CO adsorption on oxides surfaces.[28–32] This challenge arises from the inherently weak interaction between CO and oxide surfaces, making it difficult to accurately describe both CO binding energies and its vibrational frequencies. Additionally, DFT-GGA-based methods underestimate the HOMO-LUMO gap of CO by approximately 30%,[33] further limiting predictive accuracy. Collectively, these shortcomings hinder a reliable theoretical description of CO adsorption.

Moreover, standard DFT-GGA approaches face challenges in modeling oxygen vacancies in CeO$_2$.[34–37] Due to approximations in the exchange-correlation functional and the lack of proper cancellation of the Coulomb self-interaction energy, these methods tend to over-delocalize electron density. In reduced ceria, the excess electrons introduced upon reduction occupy highly localized split-off states derived from the initially empty Ce 4$f$ band, enabling the Ce$^{4+}$ → Ce$^{3+}$ reduction. Standard GGA functionals fail to capture this critical charge localization behavior. A widely adopted and computationally efficient solution is the DFT+U approach, where U represents an effective screened on-site Coulomb interaction parameter that better accounts for localized states.[38,39] Most DFT+U studies suggest that U values in the range of 4.5–6.0 eV for Ce 4$f$ states yield reliable results within the GGA framework.[40–42]

While DFT+U significantly improves the modeling of oxygen vacancies in reduced ceria, recent studies[30,31] have shown that it fails to accurately reproduce the vibrational frequencies of CO adsorbed on the three low-index surfaces of both oxidized and reduced ceria. This limitation can be addressed using hybrid-DFT approaches, which incorporate a fraction of exact Hartree–Fock exchange. Among these, the HSE06 functional[43] has proven particularly effective in capturing both the vibrational properties of adsorbed CO[30,31] and the $Ce^{4+} \rightarrow Ce^{3+}$ reduction process.[44–46] However, hybrid functionals incur significantly greater computational cost compared to DFT+U, especially when implemented within plane-wave codes.

In this study, we assess the accuracy and efficiency of the meta-GGA SCAN[47] and its numerically improved variant, r$^2$SCAN,[48] in comparison to DFT(GGA)+U and HSE06. Unlike conventional GGA functionals, SCAN and r$^2$SCAN incorporate the kinetic energy density in addition to the electron density and its gradient, allowing for a more accurate description of diverse bonding environments.[49] Our analysis focuses on two critical aspects: the accurate prediction of vibrational frequencies of CO adsorbed on ceria surfaces and the correct localization of excess charge in reduced systems.

Despite the improved computational performance of the meta-GGAs compared to hybrid-DFT, there are inherent trade-offs involved. Our results indicate that these functionals fail to adequately describe excess charge localization in reduced ceria and do not accurately capture the vibrational frequencies of CO on ceria surfaces. Although the localization issue can be remedied within a meta-GGA+U framework, the vibrational frequency discrepancies remain unresolved. We hope that our work provides a foundational basis for understanding that hybrid-DFT—or more computationally demanding approaches, such as wavefunction-based methods—are necessary for an accurate description of CO adsorption on oxides. These advanced methods promise for delivering reliable computational reference data across a wide range of well-defined oxidized and reduced oxide surfaces.

II. **MODELS AND METHODS**

Spin-polarized DFT calculations were carried out using the slab-supercell approach, with the Vienna Ab-initio Simulation Package (VASP, http://www.vasp.at; version 6.4.2).[50,51] The Ce (4f, 5s, 5p, 5d, 6s) and O (2s, 2p) electrons were treated as valence states within the frozen-core projector augmented wave (PAW) method,[52] based on the Perdew, Burke, and Ernzerhof (PBE)[27] generalized gradient approximation (GGA) functional, while the remaining electrons were considered part of the atomic cores. A plane-wave cutoff energy of 500 eV was employed. Exchange-correlation effects were described using the SCAN[47] and r$^2$SCAN[48] functionals, and results were compared against those obtained using the PBE+U methodology (with U = 4.5 eV for Ce 4$f$ states)[53] and the hybrid-DFT HSE06 functional[43]. Additionally, selected calculations applied a Hubbard U-like correction to both the SCAN and r$^2$SCAN functionals (U = 4.5 eV for Ce 4$f$ states), and the effect of dual U terms (4.5 eV for both Ce 4$f$ and O 2$p$ states) was evaluated in combination with the PBE functional.

It is important to acknowledge that dispersion forces play a role in adsorbate-surface interactions.[54,55] A previous study,[30] which examined the influence of long-range dispersion corrections in combination with PBE+U and HSE06, found negligible effects on the calculated change in the C–O bond length and the computed frequency shift ($\Delta\upsilon$) upon adsorption, apart from a slight increase ($\lesssim 0.2$ eV) in the CO binding strength to the ceria surface. Based on these findings, dispersion corrections were omitted from the present calculations.

Bulk ceria crystallizes in a cubic fluorite structure ($Fm\bar{3}m$). Its lattice constant was optimized by minimizing the stress tensor using all functionals, with a higher plane-wave cutoff energy of 650 eV, allowing the cell volume to vary during the optimization process. The Brillouin zone was sampled with a Γ-centered Monkhorst−Pack[56] (5×5×5) $k$-point mesh. During structural relaxation, total energies and atomic forces were converged within $< 1\times10^{-7}$ eV and $< |0.01|$ eV/Å, respectively.

To evaluate the accuracy and computational efficiency of the SCAN(+U) and r$^2$SCAN(+U) functionals, the low-index oxidized CeO$_2$ (111) and (110) facets, as well as the reduced CeO$_{2-x}$ (111) surface, were selected as representative models. Surface structures were simulated using slab models, adopting (1×1) surface unit cells for the oxidized surfaces and a (2×2) unit cell for the reduced CeO$_{2-x}$(111) surface (see Fig. 1). The optimized lattice parameter corresponding to each functional was used in constructing the slabs. Specifically, the (1×1) CeO$_2$ (111) surface was modeled with a four O–Ce–O trilayer-thick slab, while the (1×1) CeO$_2$ (110) surface was represented by a six CeO$_2$ layer-thick slab. For both oxidized surfaces, the Brillouin zone was sampled with a (6×6×1) *k*-point mesh for all functionals. However, in calculations employing the computational demanding HSE06 functional, a (3×3×1) *k*-point mesh was used. A vacuum space of at least 15 Å was included in all supercells to minimize spurious interactions between periodic images. The reduced CeO$_{2-x}$ (111) surface was modeled by removing a surface oxygen atom from the (111)-oriented slab with a (2×2) surface unit cell, using a (3×3×1) *k*-point mesh for Brillouin zone sampling. The localization of the excess charge corresponds to that of the most stable configuration, as previously reported.[11]

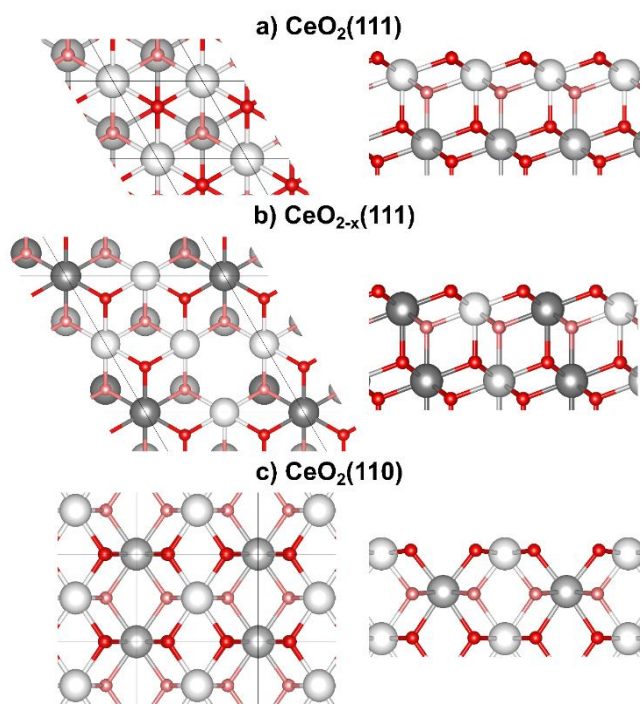

**FIG. 1.** Top and side views of the unrelaxed ceria surfaces: a) (1×1) CeO$_2$ (111), b) (2×2) CeO$_{2-x}$ (111), and c) (1×1) CeO$_2$ (110), shown here for the example of the PBE + U functional with a lattice constant of 5.46 Å lattice constant. Color code: Ce (O) atoms in the outermost layer are white (red) whereas those in deeper layers are gray (light red). The darker gray spheres indicate the positions where Ce$^{3+}$ ions localize upon lattice relaxation. This color code is consistently used in all subsequent figures.

During geometry optimization, all atoms in the three bottom atomic layers were fixed at their optimized bulk-truncated positions, while the remaining atoms were allowed to fully relax. Total energies and forces were calculated with a precision of $< 1\times10^{-8}$ eV and $< |0.001|$ eV/Å for electronic and force convergence, respectively, ensuring accuracy in both geometry optimization and vibrational frequency calculations.

The adsorption energy per CO molecule on $CeO_2$ surfaces was determined using the following equation: $E_{ads} = (E[n\cdot CO/CeO_2] - E[CeO_2] - n\cdot E[CO_{gas}])/n$, where $E[CO/CeO_2]$ is the total energy of $n$ CO molecules adsorbed on the surface, $E[CeO_2]$ is the total energy of the clean surface, and $E[CO_{gas}]$ is the energy of a CO molecule in the gas phase.

The vibrational modes of adsorbed CO were computed using a finite difference approximation of the dynamical matrix with atomic displacements of 0.015 Å, as implemented in the VASP code. Only the adsorbed CO was considered in the vibrational analysis.

For gas-phase CO, a single molecule was modeled in a cubic unit cell of $(8 \times 8 \times 8)$ Å$^3$, sampled with the Γ-point. The computed CO stretching vibrational frequencies for adsorbed CO on ceria surfaces were scaled using a method-dependent factor: $\lambda = v_{CO_{gas}}^{exp}/v_{CO_{gas}}^{calc}$ with $v_{CO_{gas}}^{exp}$ = 2143 cm$^{-1}$,[57] and calculated values of $v_{CO_{gas}}^{calc}$ = 2206, 2190, 2125 and 2233 cm$^{-1}$ for SCAN, r$^2$SCAN, PBE+U, and HSE06, respectively. The calculated C–O bond length of CO in the gas phase was 113.28, 113.48, 114.33 and 113.15 pm for SCAN(+U), r$^2$SCAN(+U), PBE(+U) and HSE06, respectively.

## III. RESULTS AND DISCUSSION

### I. Bulk CeO$_2$

#### A1. Equilibrium lattice constant

Table 1 presents the calculated equilibrium lattice constants obtained with all functionals considered in this study, compared to the experimental value of 5.41Å.[58–60] As expected, PBE exhibits the typical overestimation of approximately 1%. The application of U=4.5 eV to the Ce 4$f$ states has a minimal effect on the lattice parameter, consistent with previous studies.[46] For the SCAN and r$^2$SCAN functionals, the lattice constants are overestimated by roughly 0.2 and 0.6%, respectively; when U=4.5 eV is applied to the Ce 4$f$ states, these overestimations increase to approximately 0.6 and 0.9%. These findings agree with those of Gautam and Carter, who reported similar lattice constants using the SCAN functional both without U and with a U value of 2 eV for

the Ce 4$f$ states.[11]

It is important to note that although this U value was estimated based on the oxidation-reduction reaction enthalpy (Ce$_2$O$_3$ ⇌ CeO$_2$) and matches the value recommended for PBE+U,[46] it may not be adequate for ensuring full localization of the excess charge in reduce ceria.[40–42] Moreover, no universally optimal U value exists that accurately describes all properties of a given system.[36] Finally, the HSE06 functional yields a lattice parameter underestimated by approximately 0.4% compared to the experimental value. This result aligns with the general tendency of hybrid functionals to underestimate equilibrium volumes, as previously observed for selected metallic, semiconducting, and simple oxide systems.[61,62]

**Table 1.** Equilibrium lattice constant ($a$) and deviation from the experimental value. For the PBE+U, SCAN+U, and r$^2$SCAN+U, a U value of 4.5 eV was applied to the Ce 4$f$ states. Values shown in parentheses correspond to the results obtained with the inclusion of U.

| XC-functional | $a$ (Å) | $(a-a_{exp})/a_{exp}$ % |
|---|---|---|
| Exp.[58–60] | 5.41 | |
| HSE06 | 5.39 | −0.37 |
| PBE(+U) [a] | 5.46 (5.48) [a] | 0.92 (1.29) |
| SCAN(+U) | 5.42 (5.44) | 0.18 (0.55) |
| r$^2$SCAN(+U) | 5.44 (5.46) | 0.55 (0.92) |

[a]Using 4.5 eV for both Ce 4$f$ and O 2$p$ states yields identical results.

**A2. Electronic structure**

Figure 2 shows the total density of states (DOS) for bulk CeO$_2$ computed using the various exchange-correlation functionals considered in this study. The valence band of ceria is primarily composed of O 2$p$ states, whereas the conduction band is mainly derived from Ce 5$d$ states. The Ce 4$f$ states are located within the band gap, and their correct energetic position and degree of occupation are essential for an accurate description of both stoichiometric and reduced CeO$_2$.

In CeO$_2$, two distinct band gaps are of particular interest: (*i*) the $p$–$d$ gap ($E_g$), defined as the energy separation between the top of the O 2$p$-derived valence band and the bottom of the Ce 5$d$-derived conduction band, and (*ii*) the $p$–$f$ gap ($E_{g-f}$), which corresponds to the energy difference between the valence band maximum and the lowest unoccupied Ce 4f state. Experimentally, reported values of $E_g$ range from 5.5 to 8.0 eV [63–67], whereas $E_{g-f}$ has been measured to lie between 3.0 and 3.5 eV. [63–65] All functionals examined yield an insulating ground state, consistent with experimental observations.

The hybrid HSE06 functional predicts a p–d gap ($E_g$) of approximately 7.0 eV, which lies well within the experimental range of 5.5–8.0 eV reported by Gillen et al.[63] and is in excellent agreement with previous hybrid-DFT results for $CeO_2$.[46] The position of the Ce 4f states is also reasonably well described, with a p–f gap ($E_{g-f}$) of 3.65 eV—slightly overestimated relative to the experimental range of 3.0–3.5 eV. These results underline the reliability of the HSE06 functional for capturing the electronic structure of $CeO_2$, including both the band edges and the localization of unoccupied Ce 4f states.

In contrast, the PBE+U method with U= 4.5 eV underestimates both gaps. The calculated $E_g$ is 5.26 eV, whereas $E_{g-f}$ is approximately 2.24 eV. Da Silva et al.[46] showed that varying the U value between 0.25 and 7.25 eV, the $E_g$ gap results in a relatively stable $E_g$, whereas increasing U shifts the unoccupied Ce 4f states closer to the conduction band. Consequently. the $E_{g-f}$ gap increases from around 1.75 to 3.25 eV. These highlight the strong dependence of the Ce 4f level alignment on the selected U value.

The SCAN and r$^2$SCAN functionals yield comparable values for $E_g$ (6.04 and 6.06 eV, respectively) and $E_{g-f}$ (2.27 and 2.20 eV respectively) gaps. As with PBE+U, applying U=4.5 eV reduces $E_g$ gap by approximately 0.3 eV, whereas the $E_{g-f}$ increases by approximately 0.5 eV. For the same U=4.5 eV value, both SCAN+U and r$^2$SCAN+U functionals result in $E_g$ and $E_{g-f}$ gap values that are roughly 0.5 eV larger than those obtained with PBE+U (see Fig. 2). The $E_{g-f}$ values calculated with SCAN and SCAN+U (2.27 eV and 2.73 eV, respectively) are 0.48 and 0.80 eV larger than those reported by Gautam and Carter[68] (1.79 and 1.93 eV, respectively) using U= 2 eV. As noted above, increasing the U values will leads to a larger $E_{g-f}$ gap. No significant differences are observed between SCAN and r$^2$SCAN, or between their respective +U variants, suggesting that the primary advantage of r$^2$SCAN over SCAN lies in numerical efficiency and stability rather than in enhanced electronic structure accuracy.

In summary, both meta-GGA+U and GGA+U functionals qualitatively capture the insulating nature of $CeO_2$, but they consistently underestimate the key electronic gaps relative to experiment, with meta-GGA+U showing slightly better performance. In contrast, the HSE06 hybrid functional tends to overestimate both gaps.

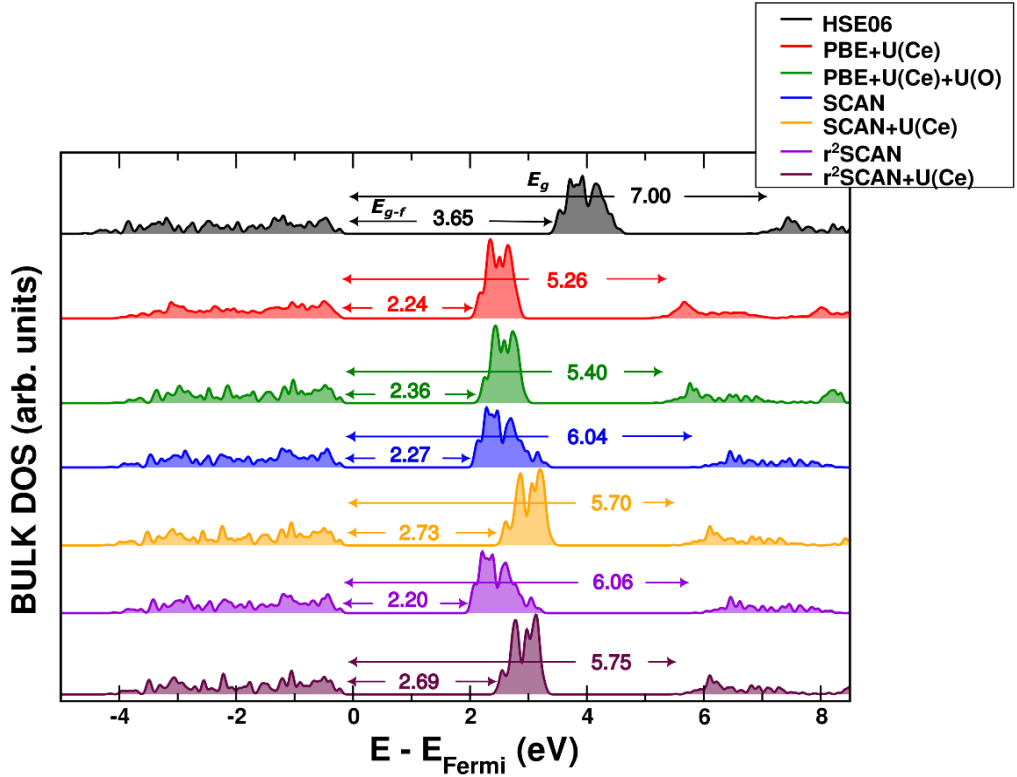

**FIG. 2.** Total density of states (DOS) for bulk $CeO_2$ bulk, calculated with different exchange-correlation functionals. All DOS plots correspond to the optimized bulk geometries. The curves are smoothed using a Gaussian level broadening of 0.05 eV. The Fermi level is set as the zero of energy. $E_g$ represents the fundamental band gap between the O 2p-derived valence band and the Ce 5d-derived conduction band (in eV). $E_{g\text{-}f}$ denotes the energy difference between the top of the valence band and the bottom of the unoccupied 4f band (in eV).

## II. Oxidized (111) and (110) surfaces

The oxidized (111) and (110) surfaces were constructed using the optimized lattice parameter corresponding to each exchange–correlation functional and subsequently relaxed. For the $CeO_2$(111) surface, we performed a benchmark study to evaluate the computational efficiency of the PBE+U, SCAN, r²SCAN, and HSE06 functionals. All calculations in this benchmark were conducted using the unrelaxed $CeO_2$(111) surface with a (1×1) unit cell, a plane-wave energy cutoff of 500 eV, a 3×3×1 Monkhorst–Pack k-point grid, and a slab model consisting of three O–Ce–O trilayers, constructed with the optimized lattice constant specific to each functional. It is important to note that the computational setup for this test differs slightly from that used in the rest of the study, particularly in terms of slab thickness and k-point sampling.

We evaluated the following performance metrics: (a) the total CPU time required for full structural relaxation; (b) the average CPU time per electronic step, based on a single-point calculation; (c) the average CPU time per ionic step, derived from the full relaxation trajectory.

All benchmarks were performed on a compute node equipped with two 56-core Intel Xeon Sapphire Rapids CPUs (totaling 112 physical cores) running at 2 GHz, on the MareNostrum 5 supercomputer hosted at the Barcelona Supercomputing Center.

The results are presented in Fig. 3. In panels (b) and (c), CPU times are normalized to the PBE+U value (set to 1.00) to emphasize relative performance differences. The data highlight the significantly higher computational cost of the HSE06 hybrid functional, which requires more than five times longer for full structural relaxation compared to PBE+U, and nearly an order of magnitude longer per electronic step. In contrast, the meta-GGA functionals SCAN and r²SCAN offer a clear computational advantage over HSE06. For single-point electronic steps, r²SCAN is slightly faster than SCAN, consistent with previous findings.[69] While both meta-GGAs require approximately twice the time per electronic step relative to PBE+U, they are still ~5 times faster than HSE06.

Interestingly, although the total CPU time for full relaxation using r²SCAN is lower than that of PBE+U, the average time per electronic and ionic step is higher. This apparent discrepancy arises because the geometry optimized in only 15 ionic steps for r²SCAN, compared to 50 for PBE+U. Thus, despite a higher per-step cost, the reduced number of optimization steps leads to a shorter overall computational time. This behavior suggests that r²SCAN achieves smoother convergence during geometry optimization—likely due to more accurate energy gradients or improved numerical stability—thereby reducing total computational effort.

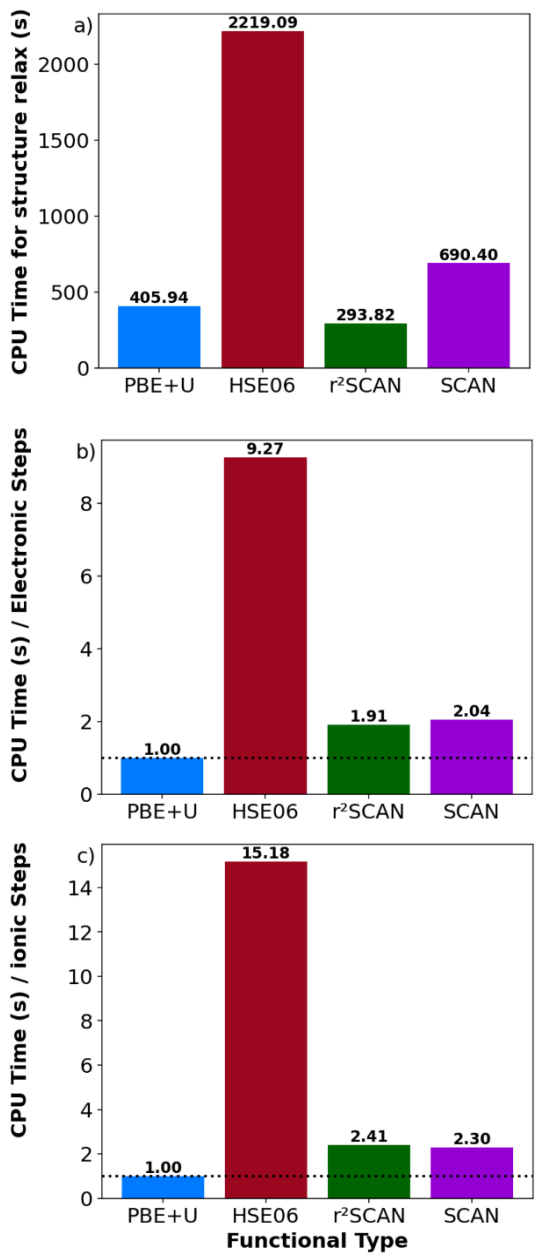

**FIG. 3.** Computational cost comparison of the exchange–correlation functionals used in this study, based on the unrelaxed $CeO_2(111)$ surface. The slab model consists of three O–Ce–O trilayers, constructed using the optimized lattice constant specific to each functional, and a 3×3×1 Monkhorst–Pack $k$-point mesh. Three performance metrics are reported: (a) total CPU time for full structural relaxation; (b) average CPU time per electronic step (from single-point calculations); and (c) average CPU time per ionic step during geometry optimization. Values in (b) and (c) are normalized to PBE+U (set to 1.00) to highlight relative performance. All calculations were performed using a plane-wave energy cutoff of 500 eV on a 112-core node of the MareNostrum 5 supercomputer.

### B1. CO adsorption on $CeO_2(111)$

Figure 4 presents the optimized structures and the calculated frequencies for CO adsorption on the oxidized CeO$_2$(111) surface at full CO coverage, using all considered exchange-correlation functionals. CO is found to adsorb atop Ce$^{4+}$ site, oriented nearly perpendicular to the surface. For both the PBE+U and HSE06 functionals, the computed CO adsorption energy ($E_{ads}$), the change in the C–O bond length ($\Delta d_{C-O}$), and the CO vibrational frequency shift, ($\Delta \upsilon$), relative to the gas-phase molecule are consistent with previous reports.[30,31] At CO saturation, the stretching vibrational frequency predicted by PBE+U is +3 cm$^{-1}$ (Figure 4), which is significantly smaller than the experimental shift of $\Delta \upsilon$= +11 cm$^{-1}$. In contrast, the hybrid HSE06 functional predicts a shift of $\Delta \upsilon$= +16 cm$^{-1}$, which is in much better agreement with the experimental result.

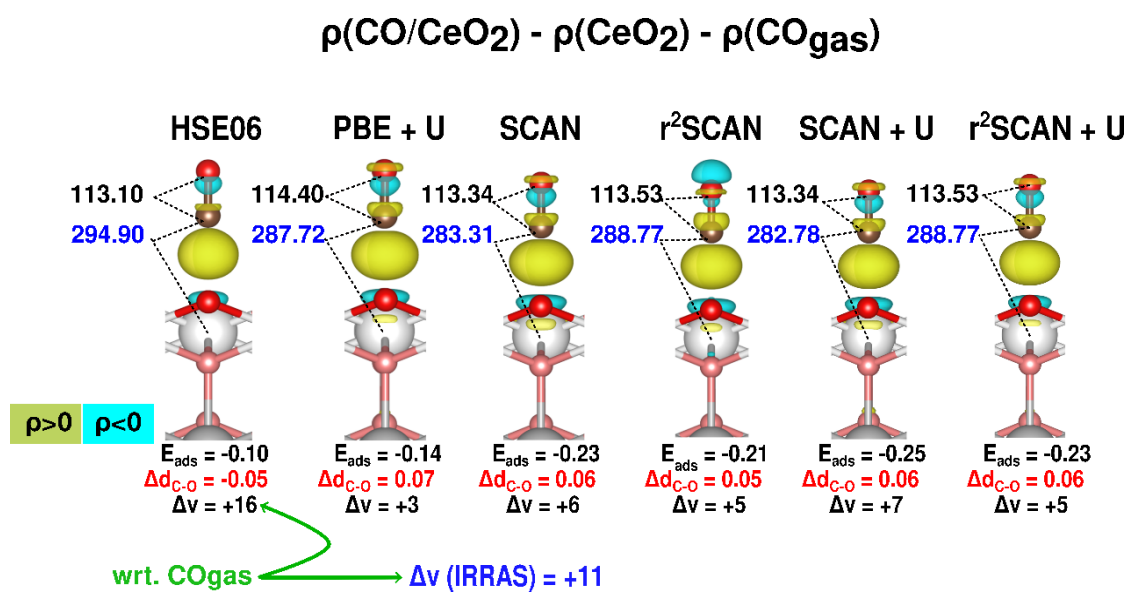

**FIG. 4.** Optimized structures and isosurfaces of the charge density difference for CO adsorption on the CeO$_2$(111) surface, computed using HSE06, PBE+U, SCAN, r$^2$SCAN, SCAN+U, and r$^2$SCAN+U functionals. The CO adsorption energy ($E_{ads}$ in eV), the change in the C–O bond length ($\Delta d_{C-O}$ in pm) relative to the corresponding gas-phase molecule for each functional and selected interatomic distances (in pm) are indicated. The CO vibrational frequency shift ($\Delta \upsilon$ in cm$^{-1}$) is reported relative to the gas-phase molecule. The experimental value of +11 cm$^{-1}$ is taken from ref.[31]

An analysis of the results obtained using the SCAN, r$^2$SCAN, SCAN+U, and r$^2$SCAN+U functionals reveals that the computed CO adsorption energies ($E_{ads}$) are approximately twice as strong compared to that predicted by the HSE06 functional, likely due to the enhanced treatment of exchange-correlation effects in the meta-GGA framework. The stretching vibrational frequency predicted by the SCAN and r$^2$SCAN approximations are +6 cm$^{-1}$ and +5 cm$^{-1}$, respectively.

Incorporating a U term results in shifts of +7 cm$^{-1}$ and +5 cm$^{-1}$, respectively. Although the meta-GGA functionals show a slight improvement in predicting the frequency shift, they still generally fall short of the accuracy achieved by the HSE06 functional. Vázquez Quesada et al.[32] reported an estimated blue-shifted vibrational frequency of +23 cm$^{-1}$, calculated at the CCSD(T)/def2-TZ/QZVPP level using a surface optimized with HSE06 functional in an embedded cluster model.

Inspection of the optimized structures for CO adsorption on the CeO$_2$(111) surface (Fig. 4) shows that the Ce–C distance predicted by the HSE06 functional is 6–12 pm longer than those obtained using the PBE+U, SCAN, r$^2$SCAN, SCAN+U, and r$^2$SCAN+U. Furthermore, the change in the C–O bond length ($\Delta d_{C-O}$) relative to the gas-phase molecule is positive ($\Delta d_{C-O} > 0$) for all functionals except HSE06, indicating that the C–O bond stretches upon adsorption by 0.05–0.07 pm. In contrast, HSE06 is the only functional for which $\Delta d_{C-O} < 0$ (−0.05 pm), meaning that the C–O bond is compressed upon adsorption. This shorter C–O bond is consistent with the larger blue shift in the CO vibrational frequency predicted by HSE06.

As previously discussed,[31] the overall change in the C–O bond upon adsorption results from synergistic charge transfer effects, specifically CO → surface σ donation and surface → CO π backdonation, which contribute to bond shortening and lengthening, respectively. Figure 5 shows the total density of states (DOS) projected onto the CO molecule for all exchange-correlation functionals, while Fig. 4 presents isosurfaces of the charge density difference (Δρ) for CO adsorption on the CeO$_2$(111) surface. Together, these figures clearly illustrate the differences in the electronic structure of the adsorbed CO as described by each functional. Compared to HSE06, all other functionals predict larger charge transfers (see Fig. 4), particularly stronger CO → surface σ donation, as evidenced by more pronounced electron density accumulation and a shorter Ce–C distance. Additionally, surface → CO π backdonation is apparent from the enhanced electron density at the O end of the elongated C–O bond, leading to $\Delta d_{C-O} > 0$.

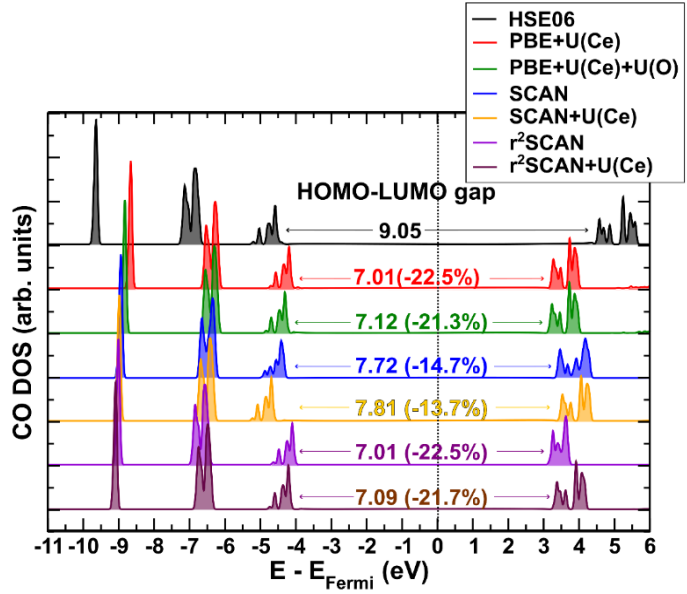

**FIG. 5.** Total density of states (DOS) projected onto the CO adsorbed atop a $Ce^{4+}$ on the (1×1) $CeO_2$(111) surface, computed using various exchange-correlation functionals. All DOS calculations correspond to the fully optimized CO/$CeO_2$(111) geometries. The curves are smoothed using a Gaussian level broadening of 0.05 eV, and the Fermi level is set as the zero of energy. Values in parentheses indicate the percentage deviation of the HOMO–LUMO gap (in eV) relative to that predicted by the HSE06 functional.

Furthermore, the previously reported underestimation of the HOMO-LUMO gap of CO by GGA functionals [33] is clearly observed in Fig. 5. Both GGA+U and meta-GGA functionals, with and without the inclusion of U, underestimate the HOMO-LUMO gap by approximately 13–23% compared to HSE06. Notably, among the meta-GGA functionals, SCAN+U shows the smallest underestimation of the HOMO-LUMO gap and yields the vibrational frequency shift that most closely matches the experimental value.

**B2. CO adsorption on $CeO_2$(110)**

For CO adsorption on the $CeO_2$(110) surface, two experimentally observed blue-shifted peaks have been reported at $\Delta\upsilon$ = +28 $cm^{-1}$ and +17 $cm^{-1}$,[31] corresponding to CO bound to $Ce^{4+}$ in atop and slightly tilted configurations, respectively. Figure 6 shows the optimized structures for both the atop and one representative tilted configuration on the (1×1) $CeO_2$(110) surface, as calculated using the HSE06 functional. The corresponding results obtained with SCAN, $r^2$SCAN, SCAN+U, and $r^2$SCAN+U functionals are summarized in Table II. The results for the HSE06 and PBE+U functionals are in line with previous studies.[30,31]

As observed for CO adsorption on the (111) surface, the computed CO adsorption energies ($E_{ads}$) are stronger using the meta-GGA functionals are stronger than those predicted by HSE06. For the atop configuration on the (110) surface, the stretching vibrational frequency shifts predicted by the SCAN and r$^2$SCAN functionals are +9 cm$^{-1}$ and +15 cm$^{-1}$, respectively. Incorporating a U term further increases the shifts to +16 cm$^{-1}$ and +18 cm$^{-1}$, respectively. Similar to the atop configuration on the (111) surface, the meta-GGA functionals offer a modest improvement in reproducing the frequency shift for the atop configuration on the (110) surface. However, they still generally fall short of the accuracy achieved by the HSE06 functional. For comparison, Herschend et al.[70] reported a blue-shifted peak by +9 cm$^{-1}$ for the top configuration obtained using hybrid-DFT embedded cluster calculations.

The discrepancies become more pronounced for the tilted configuration. In this case, only the HSE06 functional predicts a blue-shifted vibrational frequency shift, whereas all other functionals yield qualitatively incorrect red-shifted frequencies, ranging from −12 cm$^{-1}$ to −30 cm$^{-1}$. These significant deviations under the limitations of semi-local functionals in accurately describing the vibrational properties of CO adsorption on the CeO$_2$(110) surface.

The geometric and electronic structures for CO adsorption on the (110) surface were also analyzed. Table II shows that the C–Ce distances are generally shorter when using the semi-local functionals compared to the screened hybrid functional. In addition, the changes in the C–O bond length upon adsorption indicate that for the atop configuration, $\Delta d_{C–O} < 0$, meaning the bond is compressed. This compression is generally less pronounced with the semi-local functionals than with the screened hybrid functional. However, for the tilted configuration, $\Delta d_{C–O} < 0$ is observed only with the HSE06 functional, indicating a bond compression, whereas all semi-local functionals predict bond elongation ($\Delta d_{C–O} > 0$).

As with the (111) surface, the results obtained with the HSE06 functional show the best overall agreement with the experimental values.

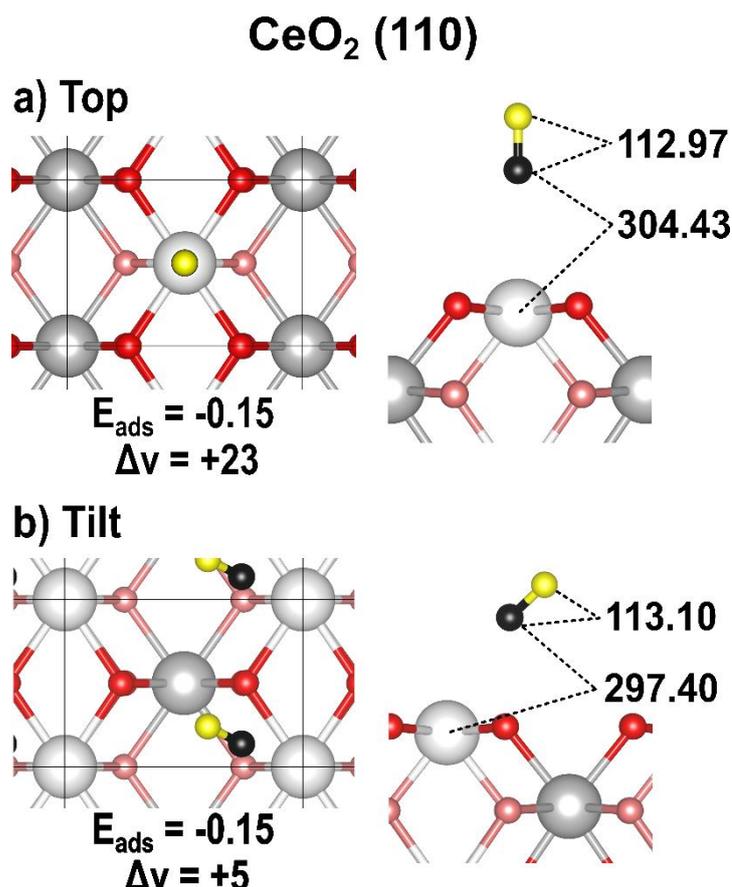

**FIG. 6.** Optimized a) atop and b) tilted structures of CO adsorption on the (1×1) CeO$_2$(110) surface, computed using HSE06 functional. The CO adsorption energy ($E_{ads}$ in eV), the change in the C–O bond length ($\Delta d_{C-O}$ in pm) relative to the corresponding gas-phase molecule for each functional and selected interatomic distances (in pm) are indicated. The CO vibrational frequency shift ($\Delta \upsilon$ in cm$^{-1}$) is reported relative to the gas-phase molecule.

**Table II.** Computed CO adsorption data on the (1×1) CeO$_2$(110) surface, computed using PBE+U, HSE06, SCAN, SCAN+U, r$^2$SCAN, and r$^2$SCAN+U functionals. The CO adsorption energy ($E_{ads}$ in eV), the change in the C–O bond length ($\Delta d_{C-O}$ in pm) relative to the corresponding gas-phase molecule for each functional and selected interatomic distances (in pm) are listed. The CO vibrational frequency shift ($\Delta \upsilon$ in cm$^{-1}$) is reported relative to the gas-phase molecule. The experimental values of +28 cm$^{-1}$ and +17 cm$^{-1}$ are taken from ref. [31].

| | CO adsorption on stoichiometric CeO$_2$(110) | | | | | |
|---|---|---|---|---|---|---|
| Site | Functional | $E_{ads}$ (eV) | $\Delta d_{C-O}$ (pm) | $d_{C-Ce}$ (pm) | $\upsilon$ (cm$^{-1}$) | $\Delta \upsilon$ (cm$^{-1}$) |
| **IRRAS experiment** | | | | | | +28/+17 |
| atop | HSE06 | -0.15 | -0.18 | 304.43 | 2166 | +23 |
| | PBE+U | -0.18 | -0.12 | 299.61 | 2154 | +11 |
| | SCAN | -0.21 | -0.13 | 294.95 | 2152 | +9 |
| | r$^2$SCAN | -0.20 | -0.13 | 300.17 | 2158 | +15 |

|  | SCAN+U | -0.24 | -0.19 | 293.49 | 2161 | **+16** |
|  | r²SCAN+U | -0.23 | -0.13 | 297.10 | 2159 | **+18** |
| tilt | HSE06 | -0.15 | -0.06 | 297.40 | 2148 | **+5** |
|  | PBE+U | -0.22 | +0.15 | 290.95 | 2127 | **-16** |
|  | SCAN | -0.27 | +0.28 | 296.57 | 2113 | **-30** |
|  | r²SCAN | -0.26 | +0.15 | 293.73 | 2131 | **-12** |
|  | SCAN+U | -0.35 | +0.22 | 291.89 | 2120 | **-23** |
|  | r²SCAN+U | -0.30 | +0.19 | 297.10 | 2124 | **-29** |

### III. Reduced (111) surface

For CO adsorption on the reduced $CeO_{2-x}(111)$ surface, a blue-shifted peak at $\Delta\upsilon = +20$ cm$^{-1}$ is experimentally observed.[31] There is a general consensus that subsurface oxygen vacancies are more stable than superficial ones and that $Ce^{3+}$ ions prefer sites away vacancies, preferably in the outermost cationic layer. In the (2×2) reduced $CeO_{2-x}(111)$ surface, the subsurface vacancy is more stable than the surface vacancy by approximately 0.4 eV (PBE+U), with $Ce^{3+}$ ions occupying next-nearest neighboring cation sites, one in the outermost cationic layer and the other one in the layer below.[11]

Before analyzing CO adsorption, we first assessed the performance of the meta-GGA functionals in describing the charge localization following oxygen vacancy formation, which drives the $Ce^{4+} \rightarrow Ce^{3+}$ reduction. Figure 7a shows isosurfaces of the spin density for a surface oxygen vacancy in $CeO_2$ (111) obtained using the $r^2SCAN$ functional, with results for SCAN being indistinguishable. The excess charge remains visibly delocalized. However, the addition of a Hubbard U-term (U = 4.5), successfully localizes the excess charge on two cations, as shown in Figures 7b and 7c. In these cases, two distinctly filled Ce $4f$ states are clearly observed below the conduction band, which is predominantly composed of Ce $5d$ and $4f$ states. This behavior closely resembles that observed in calculations using the semi-local PBE and PBE+U functionals.[36,40–42] Since the meta-GGA functionals fail to accurately capture the characteristics of the reduced surface, the addition of a U-term becomes essential when computing the vibrational frequencies of CO adsorbed on reduced ceria surfaces.

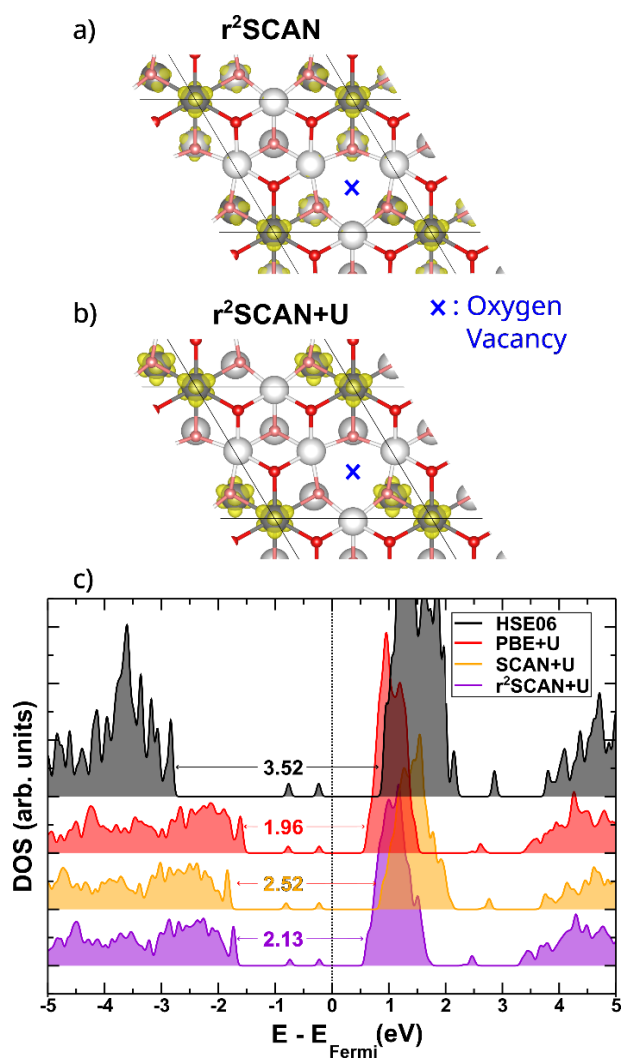

**FIG. 7.** Isosurfaces of the ferromagnetic spin density (yellow) for a surface oxygen vacancy in $CeO_2$ (111) with (2×2) periodicity, obtained using a) $r^2$SCAN and b) $r^2$SCAN+U functional. c) Total density of states (DOS) summed over spin projections (ferromagnetic state) and all atoms in $CeO_{2-x}$(111) with a surface oxygen vacancy. The DOS curves are smoothed using a Gaussian level broadening of 0.05 eV. The Fermi level is set as the zero of energy. The energy gap between the top of O 2$p$-derived valence band and the bottom of the Ce 5$d$+4$f$-derived conduction band is indicated (in eV).

In previous work, the adsorption of CO at saturation coverage on the (2×2) reduced $CeO_{2-x}$(111) surface was studied for both surface and subsurface vacancies using the HSE06 functional. Given the similarities between the calculated frequency shifts (+24 cm$^{-1}$ for a surface vacancy and 16 cm$^{-1}$ for a subsurface vacancy) and the experimental value of +20 cm$^{-1}$, distinguishing the nature of oxygen vacancies near the surface solely based on vibrational shifts remains challenging.[31] In this study, we also considered saturation coverage, as in the experiments, and focused on the case of a surface vacancy to evaluate the performance of the SCAN+U and $r^2$SCAN+U functionals. Four CO molecules were adsorbed, each positioned atop a surface cationic site (3 × $Ce^{4+}$ and 1 × $Ce^{3+}$), as shown in Figure 8. Table III presents the results for the largest frequency shift, computed using SCAN+U, $r^2$SCAN+U, and PBE+U (U = 4.5 eV), and compares them with previous HSE06 results.[31] The values obtained with PBE+U functional are consistent with earlier findings.[31]

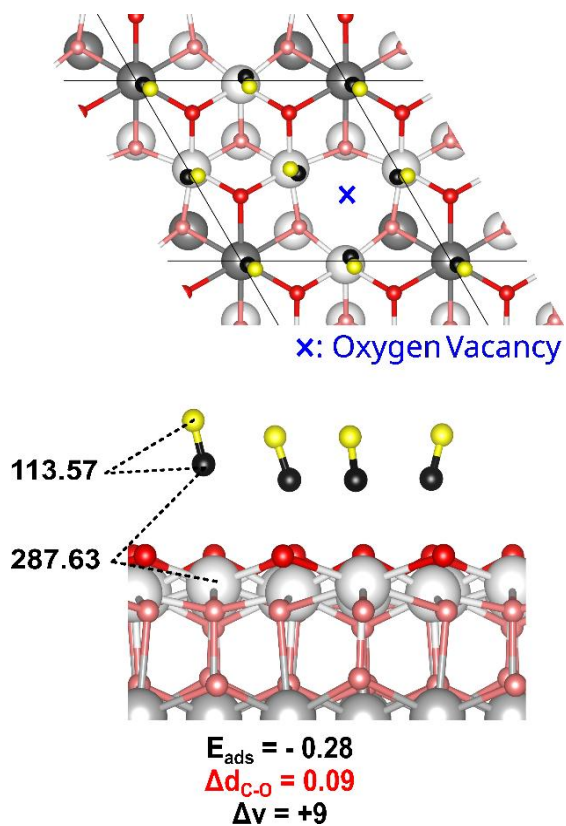

**FIG. 8.** Top and side view of optimized structure for CO adsorption at saturation coverage on the reduced (2×2) $CeO_{2-x}$(111) surface with a single surface oxygen vacancy, computed using the $r^2$SCAN+U (U = 4.5 eV). The CO adsorption energy ($E_{ads}$ in eV), the average change in the C–O bond length ($\Delta d_{C-O}$ in pm) relative to the corresponding gas-phase molecule, and average selected interatomic distances (in pm) are indicated. The CO vibrational frequency shift ($\Delta v$ in cm$^{-1}$) is reported relative to the gas-phase molecule.

The meta-GGA+U methodologies show a slight improvement in predicting the frequency shift, but they still generally fall short of the accuracy achieved by the HSE06 functional. Once again, HSE06, with a shift of +24 cm$^{-1}$, provides the closest agreement with the experimental value of +20 cm$^{-1}$.

**Table III.** Data for CO adsorption on the (2×2) $CeO_{2-x}$(111) surface at saturation coverage with a single surface oxygen vacancy, computed using PBE+U, SCAN+U, and $r^2$SCAN+U functionals. Listed values include, the CO adsorption energy ($E_{ads}$ in eV), the average change in the C–O bond length ($\Delta d_{C-O}$ in pm) relative to the corresponding gas-phase molecule for each functional, and selected interatomic distances (in pm), also presented as averages. The CO vibrational frequency shift ($\Delta v$ in cm$^{-1}$) is reported relative to the gas-phase molecule. Experimental data (+20 cm$^{-1}$) and HSE06 results are taken from ref.[31]

| | | \multicolumn{5}{c|}{CO adsorption on reduced $CeO_{2-x}$(110)} |
|---|---|---|---|---|---|---|
| Site | Functional | $E_{ads}$ (eV) | $\Delta d_{C-O}$ (pm) | $d_{C-Ce}$ (pm) | $v$ (cm$^{-1}$) | $\Delta v$ (cm$^{-1}$) |
| IRRAS experiment | | | | | | +20 |
| atop | HSE06 | -0.22 | -0.28 | 298.80 | 2167 | +24 |
| | PBE+U | -0.19 | +0.13 | 289.19 | 2148 | +5 |
| | SCAN+U | -0.30 | +0.11 | 285.19 | 2156 | +13 |
| | $r^2$SCAN+U | -0.28 | +0.09 | 287.63 | 2152 | +9 |

As discussed above for the case of CO adsorption on the oxidized $CeO_2$(111) surface, Fig. 9 shows presents the total density of states (DOS) projected onto the four CO molecules for the HSE06, PBE+U SCAN+U, and $r^2$SCAN+U functionals, now for CO adsorption on the reduced $CeO_{2-x}$(111) surface. This figure once again highlights the differences in the electronic structure of the adsorbed CO as described by each functional. In comparison to HSE06, all other functionals predict a smaller HOMO–LUMO gap. Among the meta-GGA+U functionals, SCAN+U exhibits the smallest underestimation of the HOMO–LUMO gap and also provides the vibrational frequency shift that most closely aligns with the experimental value, similar to what was observed for the oxidized surface.

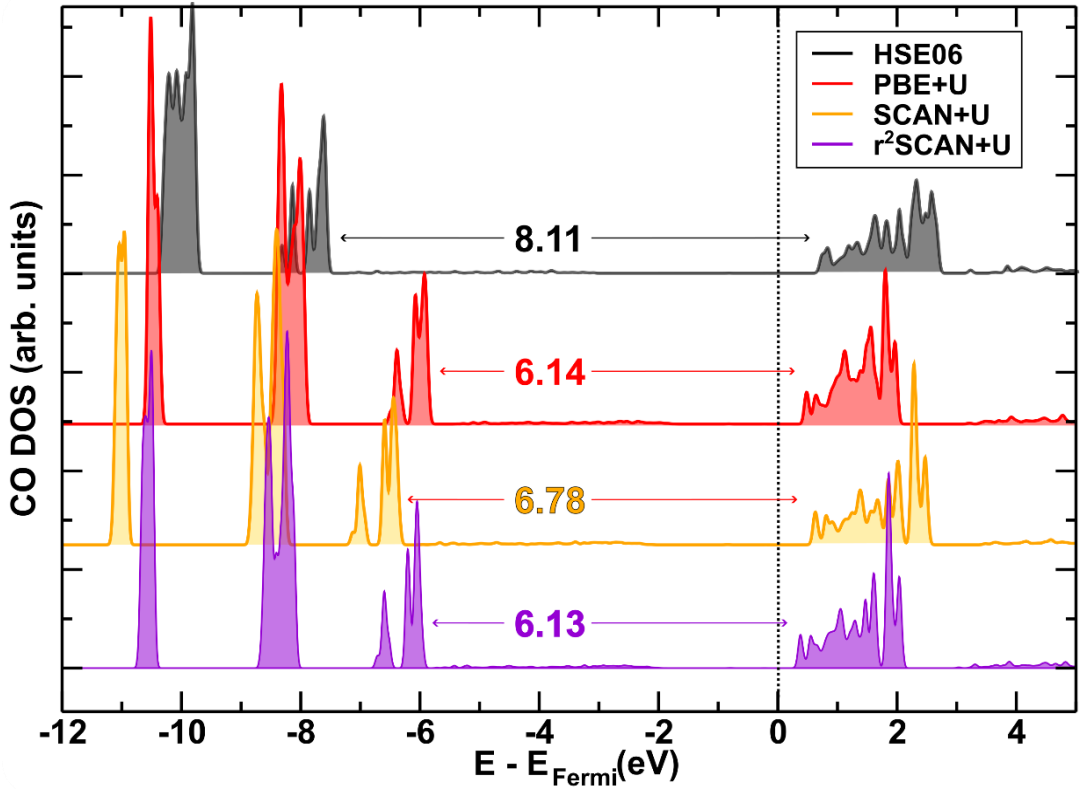

**FIG. 9.** Total density of states (DOS) projected onto 4CO adsorbed on the (2×2) $CeO_{2-x}$(111) surface, corresponding to saturation coverage with a single surface oxygen vacancy, computed using various exchange-correlation functionals. All DOS calculations correspond to the fully optimized $CO/CeO_{2-x}$(111) geometries. The curves are smoothed using a Gaussian level broadening of 0.05 eV, and the Fermi level is set as the zero of energy. The HOMO–LUMO gap (in eV) is indicated.

## IV. CONCLUSIONS

In this study, we assessed the performance of the SCAN and r²SCAN meta-GGA functionals, both with and without a Hubbard U correction, in describing the vibrational properties of CO adsorbed on low-index oxidized and reduced $CeO_2$ surfaces. Our results show that, on their own, these meta-GGA functionals fail to localize excess charge in reduced systems and do not significantly improve the accuracy of vibrational frequency predictions compared to conventional GGA+U approaches.

The inclusion of a U term partially mitigates these shortcomings by enabling charge localization. However, improvements in vibrational frequency shifts relative to PBE+U, particularly in comparison with the performance of HSE06, are less consistent. On the $CeO_2(111)$ surface, all semi-local functionals—whether or not a U term is applied—predict vibrational shifts that remain below experimental observations, although SCAN+U and r²SCAN+U show a minimal improvement over PBE+U. For the $CeO_2(110)$ surface, both SCAN+U and r²SCAN+U yield slightly improved frequency shifts for the atop CO configuration. Nonetheless, none of the semi-local functionals are able to reproduce the experimentally observed blue shift for tilted CO configurations—an effect that is correctly captured only by the hybrid HSE06 functional.

While some of these trends may have been expected given the known strengths and limitations of various functional classes, to our knowledge, this is the first systematic investigation comparing SCAN, r²SCAN, their +U variants, and HSE06 in the context of CO vibrational spectroscopy on both oxidized and reduced ceria surfaces. Our results confirm and quantify the limitations of meta-GGA functionals in these systems and provide new insight into their relative performance across different surface terminations and adsorption environments.

In conclusion, although meta-GGA+U functionals can offer a computationally affordable alternative to hybrid functionals in selected cases, they fall short of delivering the level of accuracy required for a fully reliable theoretical interpretation of experimental vibrational spectra of CO on ceria surfaces. We anticipate that this conclusion can be extended to oxide surfaces in general. These findings reaffirm the continued importance of hybrid DFT methods, such as HSE06—or even more advanced electronic structure approaches—when aiming for a quantitatively accurate and transferable description of CO adsorption on oxide-based catalytic materials.

**ACKNOWLEDGMENTS**

We acknowledge support from Grant PID2021-128915NB-I00, funded by MCIN/AEI/10.13039/501100011033 and co-financed by ERDF, EU. A.C.P. gratefully acknowledges the support of the Ramón Areces Foundation for the 2024 doctoral thesis grant in Life and Material Sciences. We also appreciate the computational resources provided by the Red Española de Supercomputación (RES) at the MareNostrum 5 (BSC, Barcelona) node.

# AUTHOR DECLARATIONS

**Conflict of Interest**

The authors have no conflicts to disclose.

**Author Contributions**

**Alexander Contreras-Payares:** Investigation (equal); Writing – review & editing (equal). **Pablo G. Lustemberg:** Investigation (equal); Supervision theoretical work (equal); Writing – review & editing (equal). **M. Verónica Ganduglia-Pirovano**: Conceptualization (lead); Supervision theoretical work (equal); Writing – original draft (lead); Writing – review & editing (equal).

# DATA AVAILABILITY

The DFT data that support the findings of this study are available in Materials Cloud {https://www.materialscloud.org/home} with the identifier DOI: …